\documentclass[%
reprint,
 amsmath,amssymb,
prb,
]{revtex4-2}

\usepackage{graphicx}% Include figure files
\usepackage{dcolumn}% Align table columns on decimal point
\usepackage[utf8]{inputenc}
\begin{document}

\preprint{APS/123-QED}

\title{A gate-tunable ambipolar quantum phase transition in a topological excitonic insulator}% Force line breaks with \\
%\thanks{A footnote to the article title}%

\author{Yande Que\(^1\)}
\author{Yang-Hao Chan\(^{2,3}\)}
\author{Junxiang Jia\(^1\)}
\author{Anirban Das\(^{4,5}\)}
\author{Zhengjue Tong\(^1\)}
\author{Yu-Tzu Chang\(^2\)}
\author{Zhenhao Cui\(^1\)}
\author{Amit Kumar\(^1\)}
\author{Gagandeep Singh\(^1\)}
\author{Shantanu Mukherjee\(^{4,5,6}\)}
\author{Hsin Lin\(^7\)}
\author{and Bent Weber\(^1\)}

\email{b.weber@ntu.edu.sg}

\affiliation{
 \(^1\)School of Physical and Mathematical Sciences, Nanyang Technological University, Singapore 637371, Singapore\\
 \(^2\)Institute of Atomic and Molecular Sciences, Academia Sinica, Taipei 106319, Taiwan\\
 \(^3\)Physics Division, National Center of Theoretical Physics, Taipei 10617, Taiwan\\
 \(^4\)Department of Physics, Indian Institute of Technology Madras, Chennai, Tamil Nadu 600036, India\\
 \(^5\)Center for Atomistic Modelling and Materials Design, Indian Institute of Technology Madras, Chennai, Tamil Nadu 600036, India\\
 \(^6\)Quantum Centre for Diamond and Emergent Materials, Indian Institute of Technology Madras, Chennai, Tamil Nadu 600036, India\\
 \(^7\)Institute of Physics, Academia Sinica, Taipei 115201, Taiwan
}

\date{\today}% It is always \today, today,
             %  but any date may be explicitly specified

\begin{abstract}
Coulomb interactions among electrons and holes in two-dimensional (2D) semimetals with overlapping valence and conduction bands can give rise to a correlated insulating ground state via exciton formation and condensation. One candidate material in which such excitonic state uniquely combines with non-trivial band topology are atomic monolayers of tungsten ditelluride (WTe\textsubscript{2}), in which a 2D topological excitonic insulator (2D TEI) forms. However, the detailed mechanism of the 2D bulk gap formation in WTe\textsubscript{2}, in  particular with regard to the role of Coulomb interactions, has remained a subject of ongoing debate. Here, we show that WTe\textsubscript{2} is susceptible to a gate-tunable quantum phase transition, evident from an abrupt  collapse of its 2D bulk energy gap upon ambipolar field-effect doping. Such gate tunability of a 2D TEI, into either \textit{n}- and \textit{p}-type semimetals, promises novel handles of control over non-trivial 2D superconductivity with excitonic pairing.
\end{abstract}

\maketitle

%\tableofcontents

\section{\label{sec:level1}Introduction}
An excitonic insulator is a correlated insulator that arises from electron-hole interactions in semimetals with overlapping electron and hole pockets at the Fermi level (\(E\textsubscript{F}\)). First proposed in the 1960s \cite{Mott1961transition, Jérome1967Excitonic, Kohn1970Two, Halperin1968Possible}, only recently has experimental evidence for the excitonic insulating state been put forth in a small number of two-dimensional (2D) electronic systems \cite{Gu2022Dipolar, Zhang2022Correlated, Chen2022Excitonic, Wang2019Evidence, Du2017Evidence, wakisaka2009excitonic, lu2017zero, Monney2010Probing, Kogar2017Signatures, Ma2022Ta2NiSe5:, Jia2022Evidence, Sun2022Evidence, gao2023evidence}. Among them, excitonic pairing has been proposed \cite{Jia2022Evidence, Sun2022Evidence, Varsano2020monolayer, Kwan2021Theory} as a contributing mechanism to stabilize the bulk energy gap in WTe\textsubscript{2} monolayers, previously understood as a single-particle band gap in the quantum spin hall (QSH) state \cite{Qi2010, Lodge2021, Zheng2016On, Tang2017Quantum}.

WTe\textsubscript{2} crystallizes in a monoclinic 1T’ lattice structure (\textbf{Figure} 1a), result of a spontaneous lattice distortion that gives rise to a doubling of the crystal unit cell and inversion \cite{ muechler2016topological, Qian2014Quantum} of the Te-\textit{p} and W-\textit{d} bands. Spin orbit coupling (SOC) \cite{Qian2014Quantum, zhao2021determination, lau2019influence, wang2023breakdown} lifts the degeneracy at the band crossing points and -- depending on the strength of the SOC -- has been predicted to give rise to a fully gapped band-insulator \cite{Zheng2016On}, or a 2D semimetal with overlapping electron and hole pockets \cite{Qian2014Quantum} (\textbf{Figure} 1b). Qian et al. \cite{Qian2014Quantum} first noted that in WTe\textsubscript{2} the SOC strength may be insufficient to open a gap, leaving it semimetallic. In contrast, a single-particle band gap of \(>\)100 meV was predicted by Zheng et al \cite{Zheng2016On} who took both SOC and exchange correlations into account. Measurement of the WTe\textsubscript{2} electronic band structure in angle resolved photoemission spectroscopy (ARPES) \cite{Tang2017Quantum}, scanning tunnelling microscopy/spectroscopy (STM/STS) \cite{Tang2017Quantum, Lüpke2020Proximity-induced}, and transport \cite{Wu2018Observation, Fei2017Edge} have confirmed the presence of a 2D bulk gap, which has since been shown to be sensitive to external electric fields \cite{Maximenko2022Nanoscale} and strain \cite{Zhao2020Strain}. 

Meanwhile, early studies reported a semimetallic bulk \cite{Jia2017prb, Song2018Observation} in heavily \textit{n}-doped WTe\textsubscript{2} monolayers, which appeared to be at odds with a band-insulator picture. A negative band gap with overlapping electron and hole pockets was found, as inferred from quasiparticle interference experiments. The presence of a long-range disorder induced Coulomb gap at the Fermi level confirmed the presence of a semimetallic state at high doping. Indeed, if a negative band gap exists in WTe\textsubscript{2}, 2D correlations would be expected at low temperature, arising from Coulomb interactions among momentum-separated electrons and holes at \(E\textsubscript{F}\). These can give rise to bound electron-hole pairs (excitons) and their condensation \cite{Jérome1967Excitonic, Varsano2020monolayer}, as illustrated in \textbf{Figure} 1b. The resulting interaction-stabilized quasiparticle gap would thus be expected to depend strongly on the interaction strength, susceptible to temperature \cite{gao2023evidence}, electric fields \cite{He2021Tunneling-tip-induced} or charge doping \cite{lu2017zero, fukutani2019electrical, chen2020doping}.

Here, we provide direct evidence of such excitonic insulating state in WTe\textsubscript{2}---for the first time demonstrating a gate-controlled quantum phase transition (QPT), as evident from a rapid collapse of the 2D bulk gap upon ambipolar field-effect doping away from the charge-neutral point (CNP).

%%%% Figure 1 %%%%%%%%%
\begin{figure*}[htbp]
\includegraphics{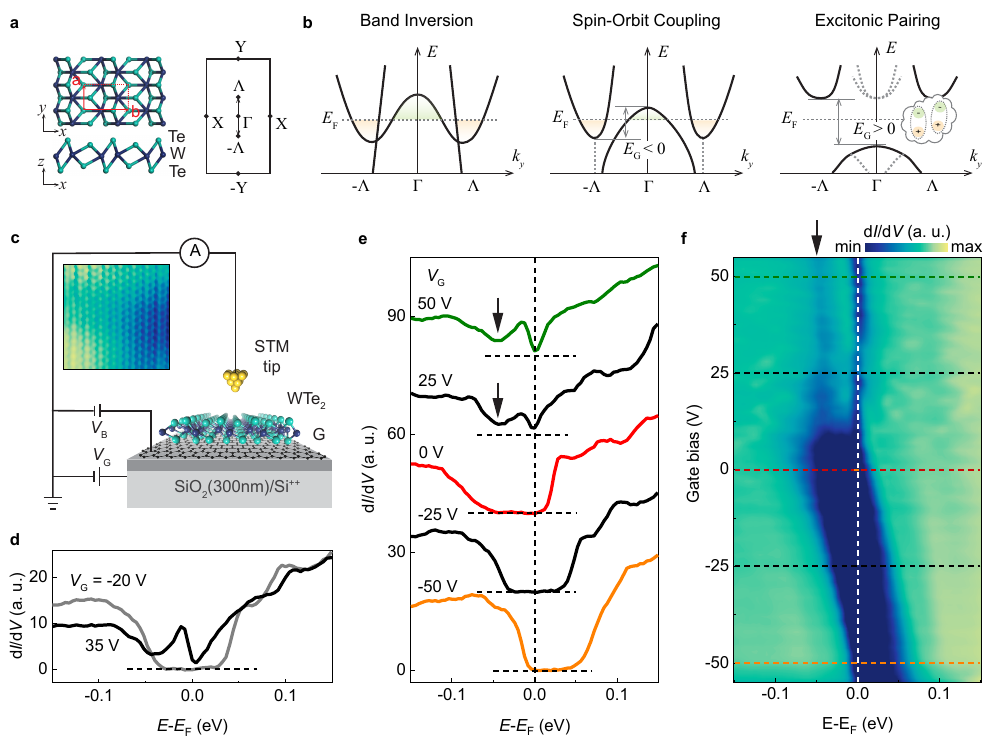}
\centering
\label{fig: fig1}
\caption{A Gate-tunable quantum phase transition (QPT) in WTe\textsubscript{2}. \textbf{a}, Atomic structure of WTe\textsubscript{2} and corresponding first Brillouin zone (BZ). The lattice constants \(a\) and \(b\) are indicated. The high symmetry points (\(\Gamma\), X, Y) and \(\Lambda\) are indicated. \textbf{b}, Band diagrams showing the formation of the 2D TEI bulk energy gap. Band inversion and SOC leave a negative bulk gap (\(E\textsubscript{G} < 0 \)), while excitonic interactions open a positive gap (\(E\textsubscript{G} > 0 \)) at low temperature. \textbf{c}, Schematic of the gated WTe\textsubscript{2}/graphene sample for STM/STS. Grounding and bias of graphene and the back-gate electrodes are indicated. The inset shows a topographic image (5 × 5 nm\(^2\)) of the WTe\textsubscript{2} surface. Long-range modulations in the measured \textit{z}-height are due to surface roughness of the underlying SiO\textsubscript{2} substrate. \textbf{d},\textbf{e}, Differential conductance (d\textit{I}/d\textit{V}) measured in the bulk of a WTe\textsubscript{2} crystal at different gate voltages ranging between -55 V and +55 V as indicated. The spectra in \textbf{e} are vertically offset for clarity, with the zero-conductance levels indicated by horizontal dashed lines. \textbf{f}, Corresponding d\textit{I}/d\textit{V} intensity map as functions of energy and gate bias from which the data in \textbf{d} and \textbf{e} were extracted (horizontal dashed lines). The Fermi level (\(E = E\textsubscript{F}\)) is indicated by the vertical dashed line in \textbf{d} and \textbf{e}. The black arrows in \textbf{e} and \textbf{f} highlight the position of the charge neutral point (CNP).}
\end{figure*}
%%%%%%%%%%%%%%%%%%%%%%%%%%%%%%%%

\section{Results and Discussions}
Our main evidence for the gate-controlled QPT is summarized in \textbf{Figure} 1d-e. We employ a field-effect gated monolayer graphene as a substrate for van-der Waals (vdW) epitaxy of WTe\textsubscript{2} (\textbf{Figure} 1c). For this, a graphene monolayer was first mechanically exfoliated onto a SiO\textsubscript{2}(300nm)/Si wafer (\textbf{Figure} S1 of the supporting information) on which WTe\textsubscript{2} was subsequently grown by vdW epitaxy in ultra-high vacuum. When a gate bias is applied to the highly  \textit{p}-doped silicon back-gate, it allows to tune the carrier density in the graphene within a range of -5.6 × 10\(^{12}\) cm\(^{-2}\) (holes) to 2.3 × 10\(^{12}\) cm\(^{-2}\) (electrons) (see \textbf{Figure} S2 of the supporting information). The associated mismatch in eletrochemical potential then allows charge transfer doping of the WTe\textsubscript{2} across the vdW gap.

Measurements of the WTe\textsubscript{2} bulk LDOS at different gate voltages (\(V\textsubscript{G}\)) are shown in \textbf{Figure} 1d-f. The QPT becomes evident as a sharp transition between an insulating (gapped) and a semimetallic (ungapped) bulk LDOS with a sudden collapse of the gap at \(V\textsubscript{G}\) = 15 V. Such collapse of a well-formed and stable 2D bulk gap cannot be explained from a single-particle picture and constitutes key-evidence for the presence of a 2D correlated insulating state. Beyond the critical voltage of the QPT (\(V\textsubscript{G} > 15\) V), only a V-shaped suppression of the LDOS remains at \(E\textsubscript{F}\), previously attributed to a Coulomb gap \cite{Song2018Observation}. A local minimum at \(E - E\textsubscript{F} = -40\) meV reflects the position of the CNP \cite{Song2018Observation}, and further confirms the presence of an \textit{n}-type semimetal. A weak effect of the gate on charge doping in the semimetallic phase over tens of volts in gate bias finally confirms a large density of states at \(E\textsubscript{F}\) in which the large carrier density screens any electric field applied. 

%%%% Figure 2 %%%%%%%%%
\begin{figure*}[htbp]
\includegraphics{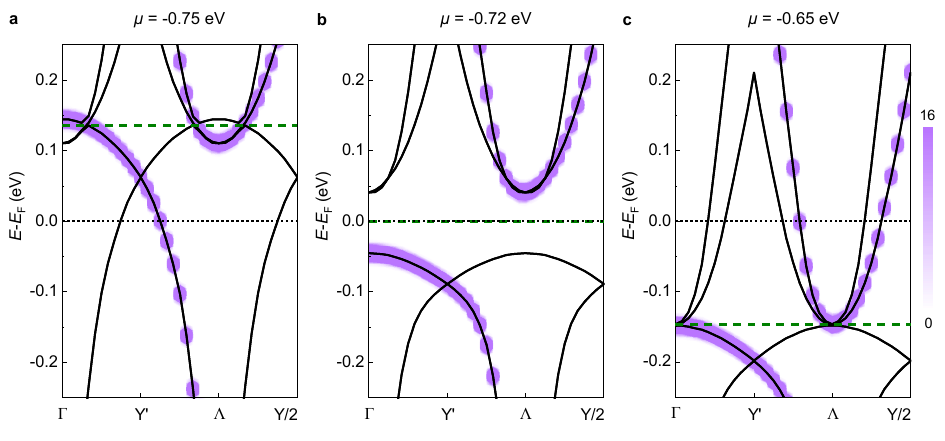}
\centering
\label{fig: fig2}
\caption{Effect of charge doping on the WTe\textsubscript{2} band structure. \textbf{a}-\textbf{c}, \(\textbf{\textit{k}}\cdot\textbf{\textit{p}}\) band structure calculation of a (\textbf{a})  \textit{p}-doped (\(\mu\) = -0.75 eV), (\textbf{b}) charge neutral (\(\mu\) = -0.72 eV), and (\textbf{c}) \textit{n}-doped (\(\mu\) = -0.65 eV) WTe\textsubscript{2} monolayer (see main text for detail) in a \(6\times1\) supercell geometry. The spectral weight of the unfolded bands are superimposed, highlighted by purple markers. Black and green dashed lines indicate the position of Fermi level and charge neutral point, respectively.}
\end{figure*}
%%%%%%%%%%%%%%%%%%%%%%%%%%%%%%%%

To gain further insight into the effects of field-effect doping, we describe the electronic structure of a WTe\textsubscript{2} monolayer in a \(\textbf{\textit{k}}\cdot\textbf{\textit{p}}\) model \cite{Jia2022Evidence, Kwan2021Theory} (See Appendix for detail). Near the \(\Gamma\) point, the Hamiltonian of a four-band model reads,
\begin{equation} \label{eq: 1}
    \hat h\left( k \right) = {\varepsilon _ + }\left( k \right) + \left[ {{\varepsilon _ - }\left( k \right) + \delta } \right]{\tau ^z} + {v_x}{k_x}{\tau ^x}{s^y} + {v_y}{k_y}{\tau ^y}{s^0}
\end{equation}
 where the \(\tau^\mu\) and \(s^\mu\)  are Pauli matrices, representing the orbital and spin degrees of freedom. \(\tau^{z} = \pm1\) refers to \textit{d} and \textit{p} orbitals, respectively. For the non-interacting Hamiltonian, we choose the same parameters as proposed in Ref. \cite{Kwan2021Theory}. We consider an interacting Hamiltonian, 
\begin{equation} \label{eq: 2}
    H_{int}  = \frac{1}{2N_k}\sum_{k,p,q,\alpha ,\beta} U(q)c_{k + q,\alpha }^+ c_{p - q,\beta }^+ c_{p,\beta }c_{k,\alpha}
\end{equation}
where \(c_{k,\alpha }^ + \) (\(c_{k,\alpha }\)) are the creation (annihilation) operators for electrons with momentum \textbf{\textit{k}} and orbital index \(\alpha\), \(N_k\) is the total number of \textit{k}-points, and \( U\left( \textbf{\textit{q}} \right) = \frac{{2{U_0}}}{{q\xi }}\tanh \frac{{q\xi }}{2}\) is a model screened interaction with a screening length \(\xi  = 25\) nm. We have adjusted the value of \({U_0}\) such that the gap after self-consistent mean-field calculations matches that observed in the experiments (\(\sim60-80\) meV). The so obtained \(U_0=25\) eV suggests \cite{Sun2022Evidence} that the system is in a spin density wave (SDW) phase with finite SDW order parameters but close to the insulator boundary. To simulate charge doping, we further introduce a chemical potential term \(\mu\) in the non-interacting part of the Hamiltonian. The folded band structure, together with the unfolded spectral weights in the BZ, are shown for different doping levels in \textbf{Figure} 2, where we clearly see the transition from an insulating (gapped)  to \textit{n}- and  \textit{p}-type semimetallic (ungapped) phase, respectively.

%%%% Figure 3 %%%%%%%%%
\begin{figure*}[htbp]
\includegraphics{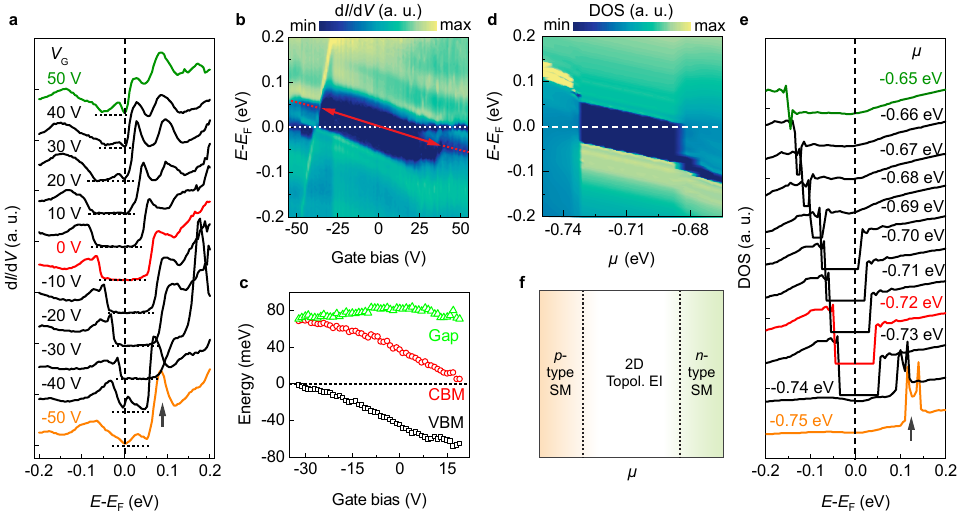}
\centering
\label{fig: fig3}
\caption{Quantum phase transition (QPT) ambipolar in field-effect doping, comparing theory and experiment. \textbf{a},\textbf{b}, Measured d\textit{I}/d\textit{V} curves (\textbf{a}) and map (\textbf{b}) taken on a nominally undoped (Fermi level midgap) WTe\textsubscript{2} crystal (see \textbf{Figure} S1 for topographic images). A QPT is clearly observed at both gate bias polarity. The red dashed lines indicate the charge neutral point (CNP) in the semimetallic WTe\textsubscript{2}, and the double-headed arrow highlights the shift of CNP from  \textit{p}-type to \textit{n}-type semimetal. \textbf{c}, Extracted band edges and gap size from the data in \textbf{b} indicating a negligible field-effect on the gap magnitude. \textbf{d},\textbf{e}, Corresponding density of states (DOS) calculated based on our \(\textbf{\textit{k}}\cdot\textbf{\textit{p}}\) model. The black arrows in \textbf{a} and \textbf{e} highlight the measured and calculated DOS in \textit{p}-type semimetal. \textbf{f}, Schematic phase diagram under electrochemical potential \(\mu\) showing \textit{n}- and  \textit{p}-type semimetal (SM) and a 2D topological excitonic insulator (TEI).} 
\end{figure*}
%%%%%%%%%%%%%%%%%%%%%%%%%%%%%%%%

A direct comparison of measured and calculated LDOS is shown in \textbf{Figure} 3, demonstrating that the QPT occurs ambipolar in gate bias, as evident from the shift of CNP from -40 meV to +40 meV. At both gate bias polarities, the QPT occurs precisely when the Fermi level (\(E = E\textsubscript{F}\)) reaches either band edge, allowing for charge to be transferred into the WTe\textsubscript{2} monolayer. An interpolation between the respective positions of the CNP in the \textit{n}- and  \textit{p}-type semimetallic phases (red double-headed arrow in \textbf{Figure} 3b) shows that at net zero doping, the Fermi level aligns precisely with the position of the CNP, and the excitonic gap is centered perfectly symmetrically about \(E = E\textsubscript{F}\). We thus suspect that the QPT is driven by a break-down of the Fermi surface's nesting condition such that the bare electronic susceptibility is suppressed. We further note that the effect of a vertical electric field in our experiments is limited to a rigid band shift in the gapped phase as shown in \textbf{Figure} 3c and that no net field-effect on the gap magnitude \cite{Maximenko2022Nanoscale} was observed. Importantly, this suggests that in our samples charge transfer doping rather than the electric fields are responsible for the QPT observed. Indeed, from a measurement of the gate-dependent LDOS on bare graphene without any WTe\textsubscript{2} coverage (\textbf{Figure} S2 in the supporting information), we estimate a chemical potential shift of \(\sim\)150 meV over a gate voltage from -35 V to +25 V, similar in order compared to that measured in WTe\textsubscript{2}, and consistent with that assumed in our \(\textbf{\textit{k}}\cdot\textbf{\textit{p}}\) calculations (100 meV), confirming the field-effect doping.

%%%% Figure 4 %%%%%%%%%
\begin{figure*}[htbp]
\includegraphics{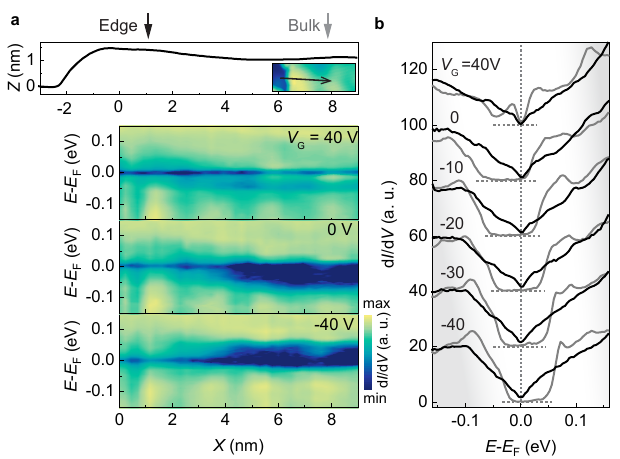}
\centering
\label{fig: fig4}
\caption{Bulk-edge correspondence in the topological excitonic insulator WTe\textsubscript{2}. \textbf{a}, Height profile and corresponding d\textit{I}/d\textit{V} intensity maps measured along a line from edge to bulk (arrow in inset) and under gate voltages from -50 V to 50 V. \textbf{b}, Individual d\textit{I}/d\textit{V} spectra comparing the bulk (gray lines) and edge (black lines) of WTe\textsubscript{2}. Spectra are vertically offset for clarity, and the horizontal and vertical dashed lines indicate the zero conductance and the Fermi level, respectively.}
\end{figure*}
%%%%%%%%%%%%%%%%%%%%%%%%%%%%%%%%

Despite the presence of Coulomb correlations, the excitonic insulator phase is expected to be topologically non-trivial as the combination of inverted bands and SOC would demand the existence of 1D metallic edge states. Indeed, real-space tight-binding calculations for WTe\textsubscript{2} ribbons with different edges reveals the edges band crossing the Fermi energy along with the gapped bulk bands (\textbf{Figure} in the supporting information). Further, as shown in \textbf{Figure} 4, we clearly observe the expected enhanced LDOS on the crystal's edges (black curves) reflecting a metallic boundary mode, regardless of the position of the Fermi level. The pseudo-gap like suppression seen in the edge state's LDOS, has previously been shown to arise from a helical Tomonaga-Luttinger liquid (TLL) ground state \cite{Jia2022Tuning,Stühler2020Tomonaga–Luttinger, Bieniek2023Theory}, and is seen to remain strictly centered at \textit{E}\textsubscript{F} regardless of doping (\textbf{Figure} 4b). From fits to a TLL model we extract a Luttinger parameter \(K\sim\) 0.3-0.4, consistent with previous work \cite{Jia2022Tuning, Bieniek2023Theory}. The metallic edge states persist even in the semimetallic phase (black curve at \(V_G = 40 V\) in \textbf{Figure} 4b), which might be expected from the ``custodial'' glide \cite{ Bieniek2023Theory, ok2019custodial} symmetry protection of the helical edge in WTe\textsubscript{2}, facilitated by a large band inversion of order hundreds of meV with band crossing and gap opening away from the high-symmetry points of the BZ.

We note that in an excitonic insulating bulk, translational symmetry is expected to be reduced \cite{Jérome1967Excitonic} due to the formation of charge density wave (CDW) order. In WTe\textsubscript{2}, a CDW with wave vector \(|\textbf{\textit{q}}\textsubscript{c}| \simeq \frac{1}{6}\frac{2\pi}{a}\) would be expected, connecting states in the hole (\(\Gamma\)) and electron (\(\pm\Lambda\)) pockets. Yet, despite claims of the excitonic insulating bulk \cite{Sun2022Evidence, Jia2022Evidence}, no CDW order has been reported to date, which has previously been explained as the contributions of entangled spin, orbital and valley degrees, paired by time-reversal symmetries, may cancel each others respective contributions \cite{Sun2022Evidence}. Breaking time reversal symmetry by application of external magnetic fields \cite{wang2023breakdown} or measurement with magnetic probes could potentially reveal the CDW (or SDW) order. Alternatively, one might expect CDW order to reemerge in the presence of translational symmetry-breaking or a spin polarization along the edge \cite{brune2012spin, Qian2014Quantum}. Indeed, we believe that the absence of translational invariance at the edge could explain periodic modulations previously observed \cite{Jia2022Tuning} in the charge density along the crystal edges of WTe\textsubscript{2} monolayers, reflecting CDW order. As we show in \textbf{Figure} S3, a fast Fourier Transform (FFT) of the energy-dependent LDOS seems to confirm this notion as the modulations have a periodicity of \(q \simeq \frac{1}{6}\frac{2\pi}{a}\), and are non-dispersive. A self-consistent real-space tight-binding calculation (see S3 of the supporting information for detail) agrees with the experimental signatures, including the period of the modulations and the exponential decay of edge state's LDOS into the 2D bulk (\textbf{Figure} S3 of the supporting information). Although atomic-level disorder at the edge can redistribute the LDOS \cite{Jia2017prb,Jia2022Tuning}, and broaden the power spectral density around \(q = k_\Lambda\), we find that the CDW modulations remain intact, and of similar frequency.

Finally, We note that the exact form of SOC in WTe\textsubscript{2} is still under debate \cite{zhao2021determination, wang2023breakdown, ok2019custodial, tan2021spin}. While an SOC type different from that assumed in Eq. (1) could influence the precise order parameters of the excitons due to the change in the bare band characters, the doping dependence and DOS inferred from our calculations should not have any significant dependence on the SOC type, as long as the resulting bare band dispersion remains similar. Future local probe spectroscopy experiments in magnetic fields and/or with magnetic probes \cite{que2023performance} should allow to to reveal the CDW, SDW or spin spiral ground states and thus shed light on the precise SOC type.

\section{Conclusion}
We have demonstrated a gate-tunable quantum phase transition (QPT), further confirming the topological excitonic insulating (TEI) state in WTe\textsubscript{2} monolayers. The QPT becomes evident from a rapid collapse of the 2D bulk gap upon ambipolar field-effect doping from a back-gate, leading to a break-down of the Fermi surface's nesting condition. The presence of a 1D metallic edge state surrounding the interaction-stabilized gapped bulk, regardless of doping, confirms that bulk-boundary correspondence persists in the 2D TEI. Periodic modulations in the LDOS at the edge, with a wave vector \(q\textsubscript{c} \simeq \frac{1}{6}\frac{2\pi}{a} = k_{\Gamma\Lambda}\), further confirm the presence of a CDW. Our work suggests WTe\textsubscript{2} and materials with similar custodial symmetry to be candidates as gate-tunable topological insulators, given the sensitive control of the bulk electronic structure arising from electron interactions and the stabilities of the edge modes at surprisingly high temperature \cite{Fei2017Edge, Bieniek2023Theory}. The interplay of topology and 2D correlated excitonic condensation might further allow to realize recent predictions of 2D triplet superconductivity with an excitonic pairing mechanism \cite{fatemi2018electrically, sajadi2018gate, Crépel2021New, crepel2022spin}.

\begin{acknowledgments}
This research is supported by National Research Foundation (NRF) Singapore, under the Competitive Research Programme “Towards On-Chip Topological Quantum Devices” (NRF-CRP21-2018-0001), with partial support from a Singapore Ministry of Education (MOE) Academic Research Fund Tier 3 grant (MOE2018-T3-1-002). H.L. acknowledges support by the Ministry of Science and Technology (MOST) in Taiwan under Grant No. MOST 109-2112-M-001-014-MY3. S.M. would like to acknowledge the new faculty seed grant from IIT Madras under project number Project No: PHY/18-19/703/NFSC/SHAA. B.W. acknowledges a Singapore National Research Foundation (NRF) Fellowship (NRF-NRFF2017-11).
\end{acknowledgments}

\appendix

\section{MBE growth}
Monolayer crystals of 1T’-WTe\textsubscript{2} were synthesized by molecular-beam epitaxy (MBE) on monolayer graphene, exfoliated on SiO\textsubscript{2} (300 nm)/Si in an Omicron Lab10 ultra-high vacuum (UHV) MBE chamber \cite{Tao2022prb, Jia2022Tuning} (base pressure below \(1 \times 10^{-10}\) mbar). The freshly exfoliated monolayer graphene substrates were annealed in UHV at 400 °C slowly (ramping rate of (\(\sim\)2 °C/min) followed by electrical contact formation by micro-soldering with indium \cite{Girit2007Soldering} in an Ar-filled glove box. Prior to MBE, monolayer graphene substrates were further degassed in UHV at 180 °C for 30 min. WTe\textsubscript{2} crystals were grown by co-deposition of W (99.998\%) and Te (99.999\%) with a flux ratio of 1:280 and substrate temperature of 160 °C for 1 hour to achieve a \(40\sim50\%\) monolayer coverage.

\section{Scanning tunnelling microscopy/spectroscopy}
Low-temperature scanning tunnelling microscopy and spectroscopy (STM/STS) were carried out in an Omicron low-temperature STM (\(\sim\)4.5 K) under UHV conditions (\(<1 \times 10^{-10}\) mbar). Chemically etched W tips or mechanically cut platinum/iridium tips were calibrated against the Au(111) Shockley surface state before spectroscopy measurements. The spectroscopy measurements were performed using standard lock-in techniques with a modulation amplitude of \(V\textsubscript{ac}\) = 2 mV and a modulation of frequency of 731.2 Hz.

\section{\(\textbf{k}\cdot\textbf{p}\) calculation}
In the four-band Hamiltonian (Eq. 1), the band energy \(\varepsilon_{\pm}(k)\) is given by
\begin{equation}
    \varepsilon_\pm(k) = \frac{1}{2}(\varepsilon_{d}(k) \pm \varepsilon_{p}(k)),
\end{equation}
where \(\varepsilon_{d}(k) = ak^2 + bk^4\) and \(\varepsilon_{p}(k) = -\frac{k^2}{2m}\) with \(a = -3\), \(b = 18\), \(m = 0.03\). The remaining parameters in Eq. 1 were set \(\nu_x = 0.5\), \(\delta = -0.9\), and \(\nu_y = 3\). The energy unit is in eV and the length unit is in Angstrom.

Decoupling the interacting Hamiltonian (Eq. 2) with the standard mean-field procedure, we get
\begin{widetext}
\begin{equation}
    H_{MF} = -\frac{1}{2N_k}[\sum_k\sum_{\alpha,\beta}\Delta_{\alpha\beta}(k, nq_c)c_{k+nq_c,\beta}^+c_{k,\alpha} + \sum_k\sum_{\alpha,\beta}\Delta_{\beta\alpha}(k, nq_c)c_{k,\alpha}^+c_{k+nq_c,\beta}],
\end{equation}
\end{widetext}
where the order parameter \(\Delta_{\alpha\beta}(k, nq_c) = \sum_qU(q)<c_{k+q,\alpha}^+c_{k+q+nq_c,\beta}>\). The numerical calculation was done in a truncated BZ in a region of \([-\frac{3}{2}q_c, \frac{3}{2}q_c] \times [-\frac{1}{4}, \frac{1}{4}]\) in units of the reciprocal lattice vector. The Hamiltonian is self-consistently solved with a uniform \(24 \times 24\) \(k\)-grid. We included upto \(n = 2\) order parameter in our calculations. The self-consistent cycle is stopped when the total energy is converged (\(< 10^{-4}\) eV). 

\bibliography{References}% Produces the bibliography via BibTeX.

\end{document}